\documentclass[titlepage,12pt,a4paper,reqno]{article}
\usepackage{amsmath,amssymb,amsfonts,graphics}
\usepackage[linktocpage=true,colorlinks,pdftex]{hyperref}
\usepackage{xcolor}
\colorlet{linkequation}{blue}
\usepackage{bm}
\usepackage{doi}
\usepackage{hyperref}
\hypersetup{colorlinks, citecolor=violet, filecolor=black, linkcolor=black, urlcolor=blue}
\newcommand*{\refeq}[1]{%
  \begingroup
    \hypersetup{
      linkcolor=linkequation,
      linkbordercolor=linkequation,
    }%
    \ref{#1}%
  \endgroup
}
\allowdisplaybreaks

\begin{document} 


\begin{titlepage}

\centerline{\LARGE \bf Gravity-mediated Dark Matter models} 
\medskip
\centerline{\LARGE \bf in the de Sitter space }
\vskip 1.5cm
\centerline{ \bf Ion V. Vancea }
\vskip 0.5cm
\centerline{\sl Grupo de F{\'{\i}}sica Te\'{o}rica e F\'{\i}sica Matem\'{a}tica}
\centerline{\sl Departamento de F\'{\i}sica}
\centerline{\sl Universidade Federal Rural do Rio de Janeiro}
\centerline{\sl Cx. Postal 23851, BR 465 Km 7, 23890-000 Serop\'{e}dica - RJ,
Brasil}
\centerline{
\texttt{\small ionvancea@ufrrj.br} 
}

\vspace{0.5cm}

\centerline{13 September 2018}

\vskip 1.4cm
\centerline{\large\bf Abstract}

In this paper, we generalize the simplified Dark Matter models with graviton mediator to the curved space-time, in particular to the de Sitter space. We obtain the generating functional of the Green's functions in the Euclidean de Sitter space for the covariant free gravitons. We determine the generating functional of the interacting theory between Dark Matter particles and the covariant gravitons. 
Also, we calculate explicitly the 2-point and 3-point interacting Green's functions for the symmetric traceless divergence-less covariant graviton. The results derived here are general in the sense that they do not depend on a particular model for the Dark Matter component and they show that the quantum effects of the interaction between the Dark Matter and the gravitons in the de Sitter space are computable by the methods of the standard Quantum Field Theory techniques.

\vskip 0.7cm 

 Dark Matter in the de Sitter space. Quantum Field Theory in curved space-time. 
\noindent

\noindent 

\end{titlepage}


\section{Introduction}

The Dark Matter (DM) paradigm has gathered observational support from different sources over the time. There are observations that estimate the velocity distribution of the galaxies in large galaxy clusters \cite{Zwicky:1937zza}, galaxy groups \cite{Klypin:1999uc} and in superclusters  \cite{Beasley:2016}. From these observations and the cluster system dynamics it is calculated that more than $90\%$ of the dynamical mass of the observed systems should be of non-luminous nature. The same mass threshold can be used to explain the velocities of the orbiting objects inside different galaxies such as the Milky Way \cite{Rubin:1970zza,Rubin:1980zd}, dwarf galaxies \cite{Bell:2000jt,Stierwalt:2017qeo} and ultra diffuse galaxies \cite{Beasley:2016}. Observations of different nature that determine the $X$-ray emissions from hot gases in equilibrium through the bremsstrahlung effect against the density dependent luminosity in $\sim 10^3$ galaxies 
\cite{Ferrari:2008jr,Vikhlinin:2005mp,Nagai:2007mt,Dietrich:2012mp} suggest that the stability of clusters in the virial model could be explained by a predominant gravitational interaction with the DM. Observations based on the gravitational lensing effect allow one to determine the cluster masses and point to a composition of more than $80\%$ DM mass in the clusters \cite{Tyson:1990yt}. The combination of the $X$-ray and gravitational lensing techniques permitted the direct observation of the separation of the stellar component from the highly excited plasma in the collision of two clusters \cite{Clowe:2006eq,Bradac:2006er,Bradac:2008eu}. Also, the DM hypothesis is consistent with the structure formation that was observed by the $2d$ Galaxy Redshift Survey and the Sloan Digital Survey and with the Lyman-alpha forest \cite{Springel:2005,Gao:2005hn} as well as with the computational simulations and the missing satellite  problem \cite{Moore:1999nt,Nierenberg:2016nri}. 
These results and other \cite{Freese:2008cz} show that the DM hypothesis is consistent with a wide spectrum of observations of the physical properties of the gravitating systems realized with different experimental techniques. However, there are no clues yet about the nature of the DM and its interaction with the Standard Model (SM) particles at the quantum scale. 

The working hypothesis is that the DM is composed of particles or fields that interact at most weakly with the components of the SM of the particle physics, that is through the three standard interactions, but quite strongly through the gravitational interaction. In order to prove this hypothesis, many experiments of direct and indirect detection of the DM particles from astrophysical and cosmical processes are undergoing. They search for the full spectrum of possible candidates as proposed by a large number of theoretical models (see for a recent review, e. g. \cite{Arun:2017}). Beside these experiments, there is the possibility of having the DM detected at the LHC \cite{Kane:2008gb} and LEP \cite{Fox:2011fx} runs. 

To describe the interaction between the DM and the SM particles, respectively, a set of phenomenological models that goes under the name of \emph{simplified models} has been proposed recently in \cite{Alves:2011wf}. In the simplified models, the envisaged interactions are described in terms of the collider physics observables: particle masses and spins, decay widths, production cross-sections and branching fractions. They are characterized by effective field Lagrangians valid at the TeV-scale that can be thought of as being derived from more fundamental theories after integrating some degrees of freedom. The simplified models are intended to help determining the sensitivity of the experimental search in colliders for the DM particles, the parametrization of possible new phenomenology and the characterization of more general models for the interaction between the DM and the SM components \cite{Alves:2011wf}\footnote{An interesting comparison of the effective field theory approach with the simplified model approach applied to the DM with scalar, spinor and vector mediators was presented in \cite{DeSimone:2016fbz}.}.

Along the line of the last application, in the present paper we are going to address the generalization of the simplified model approach to the DM with spin-2 mediating particle to a more general class of models defined in curved space-time, in particular in the de Sitter space. The main motivation behind the phenomenological study of the gravity-mediated DM interactions is the aforementioned success in observing phenomena involving the DM in the gravitating systems. Since the simplified models represent a device to describe the interactions in terms of few relevant physical parameters, it is natural to generalize them to the gravitational interactions where the same basic construction can be reproduced. This kind of models could help us understand the interaction between the DM and the space-time excitations at the quantum level. This is important in discussing phenomena where the properties of the curvature of the space-time becomes relevant as in the cosmological processes or in intense gravitational fields of massive objects. From the technical point of view, constructing these models could help us understand the theoretical tools and concepts that should be employed in analysing the quantum interaction between the DM and the space-time quantum excitations. The investigation of the DM gravitational interactions is not a new topic here. It has already been pursued in several papers from different points of view, at either macroscopic and quantum scales with massless and massive gravitons 
\cite{Tucker:1996sk,Piazza:2003ri,Parnovsky:2013rr,Noumi:2013zga,Bai:2015vca,Babichev:2016bxi,Brito:2016peb,Marzola:2017lbt,Chu:2017msm,Albornoz:2017yup}. In particular, the authors of \cite{Lee:2013bua,Lee:2014caa} proposed the simplified model for the spin-2 mediated interaction while in \cite{Kraml:2017atm} were discussed the relevant observables for the LHC running. 

This work extends the above studies by generalizing the 
simplified model for the gravity-mediated DM interaction to generally curved space-times and in particular to the de Sitter space. This space represents the simplest model of space-time that undergoes an accelerated expansion in agreement with the current astronomical observations. Therefore, understanding the interaction between the DM and the gravitons could help us understand better the observed properties of the space-time and to check the models against cosmological and intense gravitational field observations. Data could be collected much easier in these type of observations than the ones predicted by the 
next-generation DM models \cite{Bauer:2016gys}. 

In this article, we construct the gravity-mediated DM models in curved space and particularize it to the de Sitter space. Then, we show that the methods of the standard Quantum Field Theory can be used to obtain information about the interaction between DM and gravitons. In particular, we calculate the generating functional of the Green's functions in two setting: for the free gravitons and for the gravitons interacting with DM in the de Sitter space. To this end, we employ the path integral technique in the Euclidean de Sitter space which is isomorphic to the four-sphere $S^4$. The free graviton propagator has been calculated in many places in the literature in different gauges. The main contribution of this paper from this point of view is to provide the full set of Green's functions by the path integral method in the covariant gauge used previously in 
\cite{Antoniadis:1995fc,Higuchi:2001uv,Higuchi:2000ge,Allen:1986tt}. More important, we show that the interaction with the DM can be introduced as a standard interacting term and the generating functional for the interacting theory can be solved perturbatively in the coupling constant between the DM and the covariant graviton. The gravity-mediated DM models given here are independent of the particularities of the DM component which enters only through its energy-momentum tensor. Therefore, the construction can be applied to all types of DM particles.

The paper is organized as follows. In Section 2 we propose the simplified model for the DM interacting with gravitons in the curved space-time. The generalization is made by applying the same principles as in the flat space-time. In particular, the symmetries of the DM and graviton components are incorporated into the action functional. We show that the gravitational gauge symmetry of the total action imposes the vanishing of the DM energy-momentum tensor at infinity or at the boundary of the space-time manifold. In Section 3 we particularize the model to the de Sitter space. Here, we recall the basic facts about linear perturbations of the de Sitter metric and the definition of the covariant gravitons. Then we determine the generating functional of the Green's functions in the Euclidean de Sitter space obtained through a Wick transformation and we work out in detail the case of the symmetric traceless divergence-less covariant graviton. Next, we introduce the interaction with the DM and give the generating functional for the interacting theory. Also, we give the 2-point and 3-point interacting Green's functions for the symmetric traceless divergence-less covariant graviton. The results presented here show that the effects of the interaction between the DM particles and the gravitons can be computed using standard Quantum Field Theory techniques that can be applied to the simplified DM model with covariant gravitons in the de Sitter space. The last section is devoted to conclusions. 

We have collected some useful results from the literature in the Appendix. The units chosen in this paper serve the purpose of simplifying the construction. The phenomenological units can be added by dimensional analysis as usual.

\section{Gravity-mediated DM models}  

In this section, we are going to generalize the simplified DM model with covariant gravitational mediators to curved space-time. The gravitons are defined as the quantized linear perturbations of the space-time metric $g_{\mu \nu}$, that is 
\begin{align}
\tilde{g}_{\mu \nu } & = g_{\mu \nu } + h_{\mu \nu },
\label{perturb-metric-curved}
\\
\tilde{g}^{\mu \nu } & = g^{\mu \nu } - h^{\mu \nu },
\label{perturb-inverse-metric-curved}
\end{align}
where the indices are risen and lowered with the tensors $g^{\mu \nu}$ and $g_{\mu \nu}$, respectively. For example, the linear perturbation of the metric from the equation (\refeq{perturb-inverse-metric-curved}) is defined as being
\begin{equation}
h^{\mu \nu } = g^{\mu \rho} g^{\nu \sigma} h_{\rho \sigma}.
\label{h-upper-index}
\end{equation}
The generalization of the simplified DM models to curved space-times can be naturally performed by imposing the same principles as in the flat space-time. In particular, the model should contain an explicit stable DM component and should be consistent with the symmetries of all particles \cite{Alves:2011wf}. The most general action that satisfies these criteria  has the following form
\begin{equation}
\mathcal{S}_{total} = 
\mathcal{S}_{Y}
+
\mathcal{S}_{X}
+
\mathcal{S}_{int},
\label{total-action}
\end{equation}
where $\mathcal{S}_{Y}$ is the action for the free mediator in the curved space-time, $\mathcal{S}_{X}$ is the action of the DM particles and $\mathcal{S}_{int}$ is the interaction action between the mediators and the DM particles. Let us specialize the discussion to the graviton mediators. From the principle of the general covariance, the interaction action takes the following general form \cite{Lee:2013bua,Lee:2014caa}
\begin{equation}
S_{int} = - \frac{\alpha}{2} 
\int d^4 x \sqrt{-g} \, T^{\mu \nu}_{X} \, h^{Y}_{\mu \nu}, 
\label{action-XY-curved}
\end{equation}
where $\alpha$ is the coupling constant between the DM particles and the graviton. The covariant energy-momentum tensor of the DM component $T^{\mu \nu}_{X}$ depends on the unperturbed background metric $g_{\mu \nu}$. The gravitons 
$ h_{\mu \nu} $ \footnote{In what follows, we are going to drop the index $Y$ off the gravitons to simplify the notations. It is understood that they are not DM gravitons.} are unique only up to the gravitational gauge transformations
\begin{equation}
\delta_{\xi} x^{\mu } = \xi^{\mu },
\hspace{0.5cm}
\delta_{\xi} h_{\mu \nu } 
= - \mathcal{L}_{\xi } h_{\mu \nu } 
= -2 \nabla_{(\mu } \xi_{\nu) },
\label{gauge-transformations-h-curved}
\end{equation}
where $\xi^{\mu}(x)$ is of the same order as the metric perturbation $h_{\mu \nu}$. The action for the free gravitons $S_{Y} = S[h]$ is invariant under the transformations (\ref{gauge-transformations-h-curved}) as is $S_{X}$ which does not depend explicitly on $h_{\mu \nu}$.
On the other hand, since $T^{\mu \nu}_{X}$ is a rank two tensor, its variation under the gravitational gauge transformation is defined by the following relation
 \begin{equation}
\delta_{\xi} T^{\mu \nu}_{X} = - \mathcal{L}_{\xi } T^{\mu \nu}_{X}.
\label{gauge-transformation-T-curved}
\end{equation} 
Then the variation of $S_{int}$ under the transformations (\refeq{gauge-transformations-h-curved}) and (\refeq{gauge-transformation-T-curved}) takes the following form
\begin{align}
\delta_{\xi}  S_{int} =
& - \frac{\alpha}{2}
\int d^4 x \sqrt{-g} \, \nabla_{\rho} 
\left(
\xi^{\rho} 
\, 
T^{\mu \nu}_{X} h_{\mu \nu}
\right). 
\label{gauge-variation-action-curved}
\end{align}
If the space-time is bounded by the hypersurface $\Sigma$, then it follows from the Gauss' theorem that
\begin{equation}
\delta_{\xi}  S_{int} = 
- \frac{\alpha}{2}
\int_{\Sigma} d^3 x \sqrt{\gamma} \,
n_{\rho} 
\xi^{\rho} 
\, 
T^{\mu \nu}_{X} h_{\mu \nu}
,
\label{gauge-variation-action-curved-surface}
\end{equation}
where $n$ is the four-vector orthogonal to the boundary $\Sigma$
and the induced metric on the surface is 
\begin{equation}
\gamma_{ij} = \partial_i X^{\mu}\partial_j X^{\nu} g_{\mu\nu} .
\label{induced-metric-surface}
\end{equation}
Since the vector field $\xi$ is arbitrary, the invariance of the action $S_{int}$ under the gravitational gauge transformations implies the vanishing of the energy-momentum DM tensor on the boundary
\begin{equation}
T^{\mu \nu}_{X} \mid_{\Sigma} = 0.
\label{vanishing-T-DM-boundary-1}
\end{equation}
If the space-time manifold is boundless, then the invariance under the transformations (\refeq{gauge-transformations-h-curved}) and (\refeq{gauge-transformation-T-curved}) requires an asymptotic behaviour
\begin{equation}
\lim_{|\mathbf{x}| \rightarrow \infty}|T^{\mu \nu}_{X}| = 0.
\label{vanishing-T-DM-boundary-2}
\end{equation}
As usual, the norm $|\mathbf{x}|$ should be understood as calculated with a positive definite metric for which a Cauchy surface might be necessary. Note that if the gauge variations vanish on the boundary, the vanishing condition of the energy-momentum tensor are not necessary. Otherwise,
the boundary conditions (\ref{vanishing-T-DM-boundary-1}) and (\ref{vanishing-T-DM-boundary-2}) are obligatory since the simplified model method imposes that the interacting action be invariant under \emph{all} symmetries of the component fields. Whether this is a phenomenological truth or not, depends on the validity of the simplified DM model with gravity mediator at large scale.

\section{Gravitons and DM particles in de Sitter space}

The perturbations of the metric $h_{\mu \nu}$ represent the gravitons on the curved background upon quantization. The quantization procedure for linearized gravity usually involves fixing a space-time manifold (possibly by choosing a class representative) and breaking the gravitational gauge symmetry. Therefore, in what follows we are going to specialize our discussion to the de Sitter space and choose a covariant gauge in which we are going to calculate the generating functional for the graviton interacting with general DM particles.

\subsection{Covariant linear perturbations of de Sitter metric}

The metric of the de Sitter space takes different forms in different parametrizations. In particular, one can choose the global coordinates in which the line element is given by the following equation
\begin{equation}
ds^2 = - dt^2 + H^{-2} \cosh(2tH) \, d\Omega^{2}_{3},
\label{line-element-global-coordinates}
\end{equation}
where $H$ is the Hubble's constant and $d\Omega^{2}_{3}$ is the metric on the unit sphere $S^3$. Also, in order to simplify the relations, we take the natural units in which $\hbar = c = 16 \pi G_N=1$.

The derivation of the linearized free graviton action from the equation (\refeq{total-action}) is straightforward \cite{Fierz:1939ix}. By 
using the first order expansion of the metric given in the equations (\refeq{perturb-metric-curved}) and (\refeq{perturb-inverse-metric-curved}) in the Einstein-Hilbert action with a cosmological constant
\begin{equation}
S[\tilde{g}] = \int d^4 x \sqrt{-\tilde{g}} \left( \tilde{R} - 6 H^2 \right),
\label{action-full-free-gravity}
\end{equation}
where $\tilde{R}$ is the scalar curvature corresponding to the metric $\tilde{g}_{\mu \nu}$\footnote{The cosmological constant in terms of the Hubble's constant is $\Lambda = 3H^2$.}, the following Fierz-Pauli action is obtained  
\begin{align}
\mathcal{S}[h]
&=
\int d^4 x \sqrt{-g}
\left( 
\frac{1}{2}\nabla_{\mu} h^{\mu \rho} \nabla^{\nu} h_{\nu \rho}
- \frac{1}{4}
\nabla_{\mu} h_{\nu \rho}\nabla^{\mu} h^{\nu \rho} 
+ \frac{1}{4}
\left(\nabla^{\mu} h - 2\nabla^{\nu}{h^{\mu}}_{\nu}\right)
\nabla_{\mu} h 
\right)
\nonumber \\
& - \frac{1}{2}
\int d^4 x \sqrt{-g}
\left( 
h_{\mu \nu}h^{\mu \nu} + \frac{1}{2}h^2
\right)
\label{action-graviton}
\end{align}
where $h = h^{\mu}_{\mu}$ and the value of the Hubble's constant has been fixed to one for simplicity. The interaction term is given by the equation (\refeq{action-XY-curved}) while the action of the free DM particles $\mathcal{S}_{X}$ is model dependent. 

The equation of motion for the free graviton can be obtained from (\refeq{action-graviton}) by applying the variational principle
\begin{align}
{L_{\mu \nu}}^{\rho \sigma} \, h_{\rho \sigma} 
& = 
\frac{1}{2}
\left[ -\Box h_{\mu \nu} + 
\nabla_{\mu} \nabla_{\rho} {h^{\rho}}_{\nu}
+ 
\nabla_{\nu} \nabla_{\rho}{h^{\rho}}_{\mu} - \nabla_{\mu}\nabla_{\nu} h\right.  
\nonumber 
\\
&\left. 
+ g_{\mu \nu}\Box h - g_{\mu \nu}
\nabla_{\rho}\nabla_{\sigma} h^{\rho \sigma}
\right] 
+  h_{\mu \nu} + \frac{1}{2} g_{\mu \nu}h = 0.  
\label{eq-motion-graviton}
\end{align}
As can be seen from the equation above, the operator $L_{\mu \nu}^{\rho \sigma}$ is linear. 

The quantization of the field $h_{\mu \nu}$ requires fixing the gauge symmetry (\refeq{gauge-transformations-h-curved}) of $\mathcal{S}[h]$ that is also a symmetry of $\mathcal{S}_{int}$ up to a surface term, as discussed in the previous
section. The quantization in different gauges has been extensively explored in the literature. In the following, we are going to adopt the covariant gauge proposed in \cite{Antoniadis:1995fc} and used subsequently in \cite{Higuchi:2001uv,Higuchi:2000ge} to compute the graviton propagator. The gauge-fixing action has the following form
\begin{equation}
\mathcal{S}_{gf} =
-\frac{1}{4 \beta}
\int d^4 x \sqrt{-g}
\left(
2 \nabla_{\mu} h^{\mu \nu} 
- 5 \nabla^{\nu} h
\right)
\left(
2 \nabla^{\rho} h_{\rho \nu}
- 5 \nabla_{\nu} h
\right)
\, ,
\label{gauge-fixing-action}
\end{equation}
where $\beta$ is a real parameter. By applying the variational principle to the action $\mathcal{S}[h] + \mathcal{S}_{gf}$ one obtains the equations of motion in the covariant gauge
\begin{align}
{L_{\mu \nu}}^{\rho \sigma} h_{\rho \sigma}
& = 
\frac{1}{2} \Box h_{\mu \nu}
- \frac{\beta - 1}{2 \beta}
\left(
\nabla_{\mu}\nabla_{\rho} h^{\rho}_{\nu}
+ \nabla_{\nu}\nabla_{\rho} h^{\rho}_{\mu}
\right) 
+
\frac{\beta - 5}{2 \beta}
\nabla_{\mu} \nabla_{\nu} h 
\nonumber
\\
&+ 
\frac{25 - 2 \beta }{4 \beta}  g_{\mu \nu}\Box h
+ 
\frac{\beta - 3}{ \beta} 
 g_{\mu \nu} 
\nabla_{\rho}\nabla_{\sigma} h^{\rho \sigma}
- \left( 
h_{\mu \nu}+\frac{1}{2}g_{\mu \nu}h
\right) = 0.
\label{eq-motion-cov-gauge}
\end{align} 
The above equation describes the classical dynamics of the covariant linear perturbations of the de Sitter metric. Upon quantization, its solutions are the free covariant gravitons. 

\subsection{Free covariant gravitons in Euclidean de Sitter}

The free covariant gravitons were studied in 
\cite{Allen:1986tt} (see also \cite{Higuchi:2001uv,Higuchi:2000ge}). In what follows, we are going to obtain the generating functional for the Green's functions in the Euclidean de Sitter space that can be obtained by performing the global transformation 
\begin{equation}
t \rightarrow \tau =  \frac{\pi}{2} - iHt.
\label{euclidean-de-Sitter}
\end{equation}
The space obtained in this way is isomorphic to $S^{4}$
and has the metric
\begin{equation}
d s^2
= \frac{1}{H^2}
\left\{
d\tau^2 + \sin^2\tau\,
\left[
d\chi^2+\sin^2 \chi 
\left(
d\theta^2+\sin\theta^2d\phi^2
\right)
\right]
\right\}.
\label{metric-Euclidean-de-Sitter}
\end{equation}
Due to this property, the quantum excitations are defined with respect to the Euclidean vacuum. Therefore, we define the generating functional of the Euclidean Green's functions according to the general prescription as
\begin{equation}
W_{E}[\mathcal{T}] = \mathcal{N} 
\int \mathcal{D}[h]
\exp 
\left\{
i \int d^4 \bar{x}
\left[
\mathcal{L}_{E,gf} [h]
+
\mathcal{T}_{\mu \nu} (\bar{x})
h^{\mu \nu} (\bar{x} )
\right]
\right
\},
\label{generating-functional-eucl}
\end{equation}
where $\mathcal{T}$ is the source of the free graviton field. The index $E$ denotes the respective quantities on
$S^4$. The Lagrangian $\mathcal{L}_{E,gf} [h]$ corresponds to the free graviton action in the gauge from the equation (\refeq{gauge-fixing-action}). The transformation inverse to (\refeq{euclidean-de-Sitter}) leads to the Feynman propagator in the de Sitter space and the correlation functions on the de Sitter space. In order to simplify the notation, let us introduce the multi-indices 
$A = \mu \nu$, $B = \rho \sigma$ etc. Then the Green's functions generated by 
$W_{E}[\mathcal{T}]$ can be written as
\begin{equation}
G^{(n)}_{A_1 \ldots A_n} (\bar{x}_1, \cdots , \bar{x}_n )
= i^{-n} 
\frac{\delta^n W_{E}[\mathcal{T}]}{\delta \mathcal{T}_{A_1} (\bar{x}_1) \cdots \mathcal{T}_{A_n} (\bar{x}_n)}
\, .
\label{Green-functions}
\end{equation}
The generating functional $W_{E}[\mathcal{T}]$ can be obtained by applying the standard path integral techniques
(see e. g. \cite{Rivers:1987hi}). To this end, we decompose the field $h_{A}$ as follows
\begin{equation}
h_{A} (x) = h_{A,0}(x) + \chi_{A}(x),
\label{h-decomposition}
\end{equation}
where $h_{A,0}(x)$ solves the equation of motion with source
\begin{equation}
{L_{A}}^{B} h_{A,0}(x) = \mathcal{T}_{A} (x).
\label{h-equation-source}
\end{equation}
The Green's functions of the operator ${L_{A}}^{B}$ satisfy the equation 
\begin{equation}
- {L[x]}^{AB} G_{BC} (x,x') = {\delta^{A}}_{C} \, (x,x'),
\label{Green-function}
\end{equation}
where $[x]$ in ${L[x]}^{AB}$ recalls the fact that the derivatives are taken with respect to $x$. According to Green's theorem, the general solution to the equation (\refeq{h-equation-source}) is given by
\begin{equation}
h_{A,0}(x) = \int_{S^4} d^4 x' \, G_{AB} (x,x') \,
\mathcal{T}^{B} (x').
\label{h-general-solution}
\end{equation}
As reviewed in the Appendix, the scalar, vector and symmetric traceless divergence-less tensor eigenfunctions of the Laplace-Beltrami operator on the $S^4$ form a complete orthonormal set. Using the equation (\refeq{delta-function-decompose}) we can decompose the Green's functions as follows
\begin{equation}
G_{AB} (x,x') =
G_{AB}^{(s)}(x,x')
+ G_{AB}^{(v)}(x,x') 
+ G_{AB}^{(t)}(x,x')\, .
\label{Green-function-decompose}
\end{equation}
Each of these Green's functions resolve the corresponding graviton type through the equation (\refeq{h-general-solution}). Let us work out in detail the symmetric traceless divergence-less tensor $h_{A,0}(x)$. This can be calculated from the equation (\refeq{h-general-solution}) with $G_{AB}^{(t)}(x,x')$ given by the following equation
\begin{equation}
-L[x]^{AB} G_{BC}^{(t)}(x,x') = 
{\delta^{(t)A}}_{C} \, (x,x').
\label{h-tensor-equation-source}
\end{equation}
To solve the above equation for $G_{AB}^{(t)}(x,x')$, we assume that the Green's function can be decomposed in terms of the symmetric traceless divergence-less tensor eigenfunctions
\begin{equation}
G_{AB}^{(t)}(x,x') = 
\sum_{(n,a)} G_{(n,a)}^{(t)} \,
\overline{t^{(n,a)}}_{A} (x') \, t^{(n,a)}_{B} (x) ,
\label{Green-tensor-decompositon}
\end{equation} 
where the details on the tensor eigenfunctions are given in the Appendix. By substituting the equations (\refeq{Green-tensor-decompositon}) and (\refeq{delta-tensor}) in the equation (\refeq{h-tensor-equation-source}) above, one obtains the following Green's function in the mode space
\begin{equation}
G_{(n,a)}^{(t)} = \frac{2}{\lambda_{n}},
\label{Green-function-mode}
\end{equation}
where $\lambda_{n}$ is given in the Appendix. Thus, we have obtained the following results
\begin{align}
G_{AB}^{(t)}(x,x')
& = 
2 \sum_{(n,a)}  \frac{1}{\lambda_{n}} \,
 t^{(n,a)}_{A} (x)
\overline{t^{(n,a)}}_{B} (x') \, ,
\label{Green-tensor-decompositon-fin}
\\
h_{A,0}(x) & = 2 \sum_{(n,a)} \frac{1}{\lambda_{n}}
\, t^{(n,a)}_{A} (x)
\left[
\int_{S^4} d^4 x' \,
\overline{t^{(n,a)}}_{B} (x')
\mathcal{T}^{B} (x')
\right].
\label{h-general-solution-fin}
\end{align}
By substituting these results into the equation (\refeq{generating-functional-eucl}), one obtains the following generating functional for the covariant free graviton
\begin{equation}
W^{(t)}_{E}[\mathcal{T}] = 
\exp
\left\{ i \sum_{n, a}
\int_{S^4} d \bar{x} \int_{S^4} d {x'}
\,
\mathcal{T}^{A} (\bar{x}) 
\,
t^{(n,a)}_{A} (\bar{x})
\,
t^{(n,a)}_{B} (x')
\,
\mathcal{T}^{B} (x') 
\right\},
\label{generating-funct-free-graviton}
\end{equation} 
where we recognize the mode expansion of the two-point function
\begin{equation}
\mathcal{D}^{(t)}_{A B} (\bar{x}, x' )
= \sum_{n, a}
t^{(n,a)}_{A} (\bar{x})
\,
t^{(n,a)}_{B} (x').
\label{two-point}
\end{equation}
As usual, the normalization constant is defined such that
\begin{equation}
\mathcal{N}^{(t)}
\,
\int \mathcal{D}[\chi^{(t),A}]
\exp 
\left\{
i \int_{S^4} d^4 \bar{x}
\mathcal{L}_{E,gf} 
\left[
\chi^{(t)}
\right]
\right
\} = 1.
\label{normalization-constant-t}
\end{equation}
The generating functional for the scalar and vector free graviton modes can be calculated along the same line. We see that the sum over $t^{(n,a)}_{A} (\bar{x})
\,t^{(n,a)}_{B} (x')$ defines the two point correlation function  which, after the inverse Wick transformation, corresponds to the Feynman propagator of the symmetric traceless divergence-less tensor covariant graviton in the de Sitter space.

\subsection{Gravity-mediated DM models}

As we have seen above, the classical gravity-mediated DM models have the general action given by the equation (\refeq{total-action}). In order to quantize these models, we can apply the Euclidean path integral technique along the line of the previous subsection. The generating functional is then defined by the following equation
\begin{equation}
W_{int,E}[\mathcal{T}] = \mathcal{N}_{int} 
\int \mathcal{D}[h]
\exp 
\left\{
i \int_{S^4} d^4 \bar{x}
\left(
\mathcal{L}_{E,gf} [h]
+
\mathcal{L}_{E,int} [h, T^{X}]
+
\mathcal{T}_{A} (\bar{x})
h^{A} (\bar{x} )
\right)
\right
\},
\label{generating-functional-eucl-int}
\end{equation}
where $\mathcal{L}_{E,int} [h, T^{X}]$ is the Euclidean lagrangian density for the interaction between the DM particle and the graviton. By following the path integral prescription, we have
\begin{equation}
\mathcal{L}_{E,int} [h, T^{X}] = 
\frac{i \alpha}{2} 
\, 
T^{A}_{X} \frac{\delta}{\delta \mathcal{T}_{A} } .
\label{lagrangian-eucl-int}
\end{equation}
By plugging the equation (\refeq{lagrangian-eucl-int}) into the equation (\refeq{generating-functional-eucl-int}), the following expression is obtained 
\begin{align}
W_{int,E}[\mathcal{T}] & = \mathcal{N}_{int}
\exp
\left[-\frac{\alpha}{2}
\int_{S^4} d^4 x
\, 
T^{A}_{X} \frac{\delta}{\delta \mathcal{T}_{A} }
\right]
W_{E}[\mathcal{T}], 
\label{generating-functional-int-1}
\\
\mathcal{N}_{int} & = 
\exp
\left[-\frac{\alpha}{2}
\int_{S^4} d^4 x
\, 
T^{A}_{X} \frac{\delta}{\delta \mathcal{T}_{A} }
\right]
W_{E}[\mathcal{T}] \bigg\rvert_{\mathcal{T} = 0} \, .
\label{normalization-constant-int}
\end{align}
From the equations (\refeq{generating-functional-int-1}) and (\refeq{normalization-constant-int}) we can read off the Green's functions for the interaction between the DM particle and the covariant graviton
\begin{align}
\mathcal{G}^{(n)}_{A_1 \ldots A_n} (\bar{x}_1, \cdots , \bar{x}_n )
&= \mathcal{N}_{int}
\prod_{j = 1}^{n} \frac{\delta}{\delta \mathcal{T}_{A_j} (\bar{x}_j)}
\nonumber
\\
&\times
\sum_{m=1}^{\infty}
\left[
1 + \left( \frac{i \alpha}{2}\right)^{m}
\int_{S^4} \prod_{l=1}^{m} d^4 x_{l} \,
T^{A_l}_{X} (x_l) \frac{\delta}{\delta \mathcal{T}_{A_{l} (x_l )} }
\right] \, W_{E}[\mathcal{T}]\bigg\rvert_{\mathcal{T} = 0}.
\label{Green-fct-int-fin}
\end{align}
The terms in the right hand side of the above expression can be calculated from the perturbation series in $\alpha$, the coupling constant of the interaction between the gravitons and the DM particles which is expected to be small. Then the Wick's theorem can be applied order by order as in the standard Quantum Field Theory. 

We exemplify the above general results in the case of the symmetric traceless divergence-less tensor gravitons discussed in the previous subsection. There the generating functional $W^{(t)}_{int,E}[\mathcal{T}]$ given by the equations (\refeq{generating-functional-int-1}) 
with the generating functional for the free gravitons from the equation (\refeq{generating-funct-free-graviton})
takes the following form
\begin{align}
W_{int,E}[\mathcal{T}]  & = \mathcal{N}^{(t)}_{int}
\exp
\left[-\frac{\alpha}{2}
\int_{S^4} d^4 x
\, 
T^{A}_{X} \frac{\delta}{\delta \mathcal{T}_{A} }
\right]
\nonumber
\\
& \times
\exp
\left\{ i \sum_{n, a}
\int_{S^4} d \bar{x} \int_{S^4} d {x'}
\,
\mathcal{T}^{C} (\bar{x}) 
\,
t^{(n,a)}_{C} (\bar{x})
\,
t^{(n,a)}_{B} (x')
\,
\mathcal{T}^{B} (x') 
\right\}. 
\label{generating-functional-int-1-t}
\end{align}
The Green's functions are given by (\refeq{Green-fct-int-fin}) with the appropriate substitutions
\begin{align}
\mathcal{G}^{(t,n)}_{A_1 \ldots A_n} (\bar{x}_1, \cdots , \bar{x}_n )
&= \mathcal{N}^{(t)}_{int} \,
\prod_{j = 1}^{n} \frac{\delta}{\delta \mathcal{T}_{A_j} (\bar{x}_j)}
\nonumber
\\
&\times
\sum_{m=1}^{\infty}
\left[
1 + \left( \frac{i \alpha}{2}\right)^{m}
\int_{S^4} \prod_{l=1}^{m} d^4 x_{l} \,
T^{A_l}_{X} (x_l) \frac{\delta}{\delta \mathcal{T}_{A_{l} (x_l )} }
\right]
\nonumber
\\
& \times
\exp
\left\{ i \sum_{n, a}
\int_{S^4} d \bar{x} \int_{S^4} d {x'}
\,
\mathcal{T}^{C} (\bar{x}) 
\,
t^{(n,a)}_{C} (\bar{x})
\,
t^{(n,a)}_{B} (x')
\,
\mathcal{T}^{B} (x') 
\right\}
\bigg\rvert_{\mathcal{T} = 0}.
\label{Green-fct-int-fin-t}
\end{align}
At the lowest order in the coupling constant between the DM and the graviton, the formula (\refeq{Green-fct-int-fin-t}) gives the following 2-point and 3-point Green's functions
\begin{align}
\mathcal{G}^{(t,2)}_{A_1 A_2} (\bar{x}_1, \bar{x}_2 )
& \simeq
i \mathcal{N}^{(t)}_{int} \, \mathcal{D}^{(t)}_{A_1 A_2}
(\bar{x}_1, \bar{x}_2 ) \, ,
\label{Green-fct02-t-first-order}
\\
\mathcal{G}^{(t,3)}_{A_1 A_2 A_3} 
(\bar{x}_1, \bar{x}_2, \bar{x}_3)
& \simeq 
- \frac{i \alpha}{2} \mathcal{N}^{(t)}_{int} \,
\int_{S^4} d^4 x \, T^{A}_{X} (x)
\left[
\mathcal{D}^{(t)}_{A A_3}( x, \bar{x}_3 ) + 
\mathcal{D}^{(t)}_{A_3 A} (\bar{x}_3, x )
\right]
\nonumber\\
& \times
\left[
\mathcal{D}^{(t)}_{A_1 A_2} (\bar{x}_1, \bar{x}_2 )
+
\mathcal{D}^{(t)}_{A_2 A_1} (\bar{x}_2, \bar{x}_1 )
\right].
\label{Green-fct03-t-first-order}
\end{align}
This shows that the DM-graviton processes are affected by the DM energy-momentum tensor distributed throughout the full space-time. The space-time coordinate notation can be easily recovered by going from the multi-index $A$ to $\mu\nu$. For example, the equation (\refeq{Green-fct02-t-first-order}) reads in the space-time notation 
\begin{equation}
\mathcal{G}^{(t,2)}_{\mu_1 \nu_1 \, \mu_2 \nu_2} 
(\bar{x}_1, \bar{x}_2 )
 \simeq
i \mathcal{N}^{(t)}_{int} \,
\mathcal{D}^{(t)}_{\mu_1 \nu_1 \, \mu_2 \nu_2}
(\bar{x}_1, \bar{x}_2 ) \, .
\label{Green-fct02-t-first-order-sp}
\end{equation}
The concrete form of the Green's functions depend on the specific choice of the DM particles. However, the above results are general and can be applied to all DM models.

\section{Conclusions}

In this work we have generalized the simplified DM model with graviton interaction to a curved space-time background. Then, we have particularized our generalization to the phenomenologically interesting de Sitter space that represents the simplest model of the space-time that undergoes an accelerated expansion. The gravitational gauge invariance of the graviton must be a symmetry of the total action if we want our model to satisfy the usual requirements of the simplified model construction. Next, we have recalled the dynamics of the metric perturbations in the linearized gravity approach since this is the conventional framework for discussing the gravitons. In the linear gravity, we have calculated the generating functional of the Green's function for the free graviton in a covariant gauge in the Euclidean de Sitter space. Also, we have obtained the generating functional for the gravity-mediated interaction between the DM particles and the covariant gravitons and we have shown that this can be treated by standard path integral methods. And finally, 
we have calculated explicitly the 2-point and 3-point interacting Green's functions for the symmetric traceless divergence-less covariant gravitons. 

Note that our construction is model-independent in the sense that it does not depend on specific assumptions about the DM component. Therefore, it can be applied to any type of DM matter that has a calculable energy-momentum tensor. This application constitutes the object of undergoing investigations. In specific applications, the dynamics of the DM plays an important role. The construction given above can be easily extended to accommodate an explicit action for the DM particles. 

Since the technique used here is standard Quantum Field Theory, the applications of the present results to calculate quantum processes between the gravitons and the DM is immediate. That can be used to compare the predictions of these models to observations concerning the DM interactions in the cosmological phenomena in which the de Sitter structure becomes relevant.

\section*{Acknowledgements}
It is a pleasure to acknowledge M. C. Rodriguez for discussions and to an anonymous reviewer for providing insightful comments.

\section*{Appendix}

In this section, we are collecting some results that have been used throughout the paper. Most of them were presented previously in \cite{Antoniadis:1995fc,Higuchi:2001uv}.

The eigenfunctions of the Laplace-Beltrami operator on $S^4$ are classified according to their transformation properties as scalars $f^{(n,a)}$, vectors $v^{(n,a)}_{\mu}$ and symmetric tensor $t^{(n,a)}_{\mu \nu}$, where $n$ and $a$ are the labels for the spherical modes. The corresponding equations are
\begin{align}
\Box f^{(n,a)} &= - \lambda_{n} f^{(n,a)}
\, ,
\label{LB-scalar}
\\
\Box v^{(n,a)}_{\mu} &= - \left( \lambda_{n} - H^2 \right)
v^{(n,a)}_{\mu}
\,
,
\qquad
n > 1
,
\label{LB-vector}
\\
\Box t^{(n,a)}_{\mu \nu} &= - \left( \lambda_{n} - 2 H^2 \right)
t^{(n,a)}_{\mu \nu}
\,
,
\qquad
\, \, \, \,
n \geqslant 2.
\label{LB-tensor}
\end{align} 
where $\lambda_{n} = n(n+3) H^{2}$ and $n \in \mathbb{N}$. In the equation(\ref{LB-tensor}) the symmetric tensor $t^{(n,a)}_{\mu \nu}$ satisfies in addition to the eigenvalue equation, the following conditions
\begin{equation}
\nabla^{\nu} t^{(n,a)}_{\mu \nu} = 0,
\qquad
{t^{(n,a)\mu}}_{\mu} = 0 \, .
\label{LB-tensor-traceless-divergentless}
\end{equation}
The normalization conditions for the eigenfunctions are
\begin{align}
\langle
F_{1}^{(n,a)} , F_{2}^{(m,b)} 
\rangle
& = \delta^{nm} \delta^{ab} \, ,
\label{normalization-conditions}
\\
\langle
F_{1}, F_{2}   
\rangle
& = \int_{S^4} \overline{F_1} \cdot F_{2} \, , 
\label{scalar-product}
\end{align}
where $F$ stands for any of the eigenfunctions above and 
$\cdot$ denotes the vector and tensor index contraction. 
The sets of eigenfunctions are complete with respect to the scalar product defined above and the families of eigenfunctions, i. e. scalar, vector and symmetric traceless divergence-less tensor, are mutually orthogonal to each other. In general, the tensor eigenfunctions are determined by the scalar and vector eigenfunctions. 

The $\delta$-function for the symmetric tensors is defined by the following equation
\begin{equation}
\int_{S^4} d^4 x \,
\delta_{\mu \nu \rho \sigma} (x,x') t^{\rho \sigma} (x' )
=
t_{\mu\nu} (x).
\label{delta-function}
\end{equation}
Due to the orthogonality property, the $\delta$-function has the following decomposition
\begin{align}
\delta^{(s)}_{\mu \nu \rho \sigma} (x,x') & = \sum_{n,a}
\left[ \overline{f^{(n,a)}}_{\mu \nu} (x') 
\, f^{(n,a)}_{\rho \sigma} (x)
+
\overline{k^{(n,a)}}_{\mu \nu} (x') 
\, k^{(n,a)}_{\rho \sigma} (x)
\right]
,
\label{delta-scalar}
\\
\delta^{(v)}_{\mu \nu \rho \sigma}
(x,x') & = \sum_{n,a} \overline{v^{(n,a)}}_{\mu \nu} (x') \, v^{(n,a)}_{\rho \sigma} (x),
\label{delta-vector}
\\
\delta^{(t)}_{\mu \nu \rho \sigma} (x,x') & = \sum_{n,a} \overline{t^{(n,a)}}_{\mu \nu} (x') 
\, t^{(n,a)}_{\rho \sigma} (x),
\label{delta-tensor}
\end{align}
The tensor quantities constructed from the scalar and vectors are
\begin{align}
f_{\mu \nu}^{(n,a)} & = 
2 \left(
\nabla_{\mu} \nabla_{\nu}-\frac{1}{4} g_{\mu \nu}
\Box f^{(n,a)}
\right)
\left[
3\lambda_n(\lambda_n-4H^2)
\right]^{-\frac{1}{2}}
\,
,
\label{f-scalar-tensor}
\\
k_{\mu \nu}^{(n,a)} & = \frac{1}{2} g_{\mu \nu} f^{(n,a)}
\, ,
\label{k-scalar-tensor}
\\
v_{\mu \nu}^{(n,a)} & = 
2 \left( \lambda_n - 4 H^2 \right)^{-\frac{1}{2}}
\left[
\nabla_{\mu} v_{\nu}^{(n,a)} + \nabla_{\nu} v_{\mu}^{(n,a)}
\right]
\, .
\label{v-vector-tensor}
\end{align}
Using the equations (\refeq{delta-scalar}) - (\refeq{v-vector-tensor}), one can decompose the delta function from the equation (\refeq{delta-function}) as follows
\begin{equation}
\delta_{\mu \nu \rho \sigma}(x,x') = 
\delta^{(s)}_{\mu \nu \rho \sigma}(x,x')
+ \delta^{(v)}_{\mu \nu \rho \sigma}(x,x') 
+ \delta^{(t)}_{\mu \nu \rho \sigma}(x,x')\, .
\label{delta-function-decompose}
\end{equation}

\end{document}